\documentclass{kluwer}    % Specifies the document style.
\usepackage{epsfig}

\newdisplay{guess}{Conjecture}

\begin{document}                                                                
                   
\begin{article}
\begin{opening}         
\title{Objective Determination of Star Formation Histories
%\footnote{To
%appear in `Astrophysical Dynamics', a meeting in memory of Franz Kahn,
%eds David Berry, Dieter Breitschwerdt, Antonio da Costa and John
%Dyson, ApSpSci, Special Issue 1999}
}

\author{Gerard \surname{Gilmore$^1$}, Xavier \surname{Hernandez$^{1,2}$}, 
\& David \surname{Valls-Gabaud$^{1,3}$}}  
\runningauthor{Gilmore, Hernandez \& Valls-Gabaud}
\runningtitle{Objective Star Formation Histories}
\institute{$^{1}$IoA Cambridge, $^{2}$Arcetri Observatory, $^{3}$CNRS, 
Toulouse Observatory }

\begin{abstract}
Objective determination of a star formation history from a
colour-magnitude diagram, independently of assumed parametric
descriptions, is necessary to determine the evolutionary history of a
galaxy.  We introduce a new method for solving maximum likelihood
problems through variational calculus, and apply it to the case of
recovering an unknown star formation history, $SFR(t)$, from a
colour-magnitude diagram. This method provides a non-parametric
solution with the advantage of requiring no initial assumptions on the
$SFR(t)$.  As a full maximum likelihood statistical model is used, one
may exploit all the information available in the photometric data
which can be explained by available isochrones.  Extensive tests of
the method show it to be reliable under noise conditions comparable to
those appropriate for real observations. Implications for future
surveys, such as the GAIA mission are significant: accurate
determination of the evolutionary history of the Milky Way is limited
only by adequate spectroscopy and/or photometry to provide a
determination of stellar metallicity and temperature for a sufficient
sample of stars of luminosities at and above an old main-sequence
turnoff.
\end{abstract}
\end{opening}

\section{Introduction}

A primary scientific goal in the quantification of galactic evolution is
the derivation of the star formation histories, as described by the
temporal evolution of the star formation rate, $SFR(t)$.
%, and the
%cumulative numbers of stars formed, of the bulge, inner disk, solar
%neighbourhood, outer disk and halo of the Milky Way. 
 In practice,
uncertainties in the theories of stellar formation and evolution, as
well as degeneracy 
%in a stars' observational parameters 
between age
and metallicity, not to mention observational errors and unknown
distance and reddening corrections, make inferring $SFR(t)$ for mixed
stellar populations a difficult task. Even assuming a known stellar initial
mass function ($IMF$) and metallicity, a given set of isochrones and
no distance or reddening uncertainties, recovering the $SFR(t)$ which
gave rise to a given stellar mix is not trivial.

The increasing application of HST studies which resolve the stellar
populations of nearby systems has initiated quantitative investigation
of the $SFR(t)$ in these systems through the comparison of observed and
synthetic colour-magnitude (C-M) diagrams (e.g. Chiosi et al. (1989),
Aparicio et al. (1990) and Mould et al. (1997) using Magellanic and
local clusters, and Mighell \& Butcher (1992), Smecker-Hane et
al. (1994), Tolstoy (1995), Aparicio \& Gallart (1995) and Mighell
(1997) using dSph companions to the Milky Way).  These studies
construct a statistical estimator to compare a synthetic C-M
diagram constructed from an assumed $SFR(t)$ with an observed
one, and then select the $SFR(t)$, from amongst a set of plausible
ones, which maximizes the value of this estimator (e.g. Tolstoy \&
Saha 1996).  The most rigorous estimator is probably the likelihood,
as defined through Bayes's theorem. In practice this states that one
should look for the model which maximizes the probability of the
observed data set having arisen from it. The apparent robustness of
the approach is undermined by the iagramsree of subjectivity associated
with defining the set of plausible models one is able to
consider. Further, as none of the statistical estimators has an
absolute normalization, in the end one is left with that model, of the
ones one started by proposing, which best reproduces the data. This 
best model is not necessarily a ``good'' approximation to the true
$SFR(t)$. The likelihood of the data having arisen from a particular
model can only be calculated if one has the data, the errors, and the
particular model fully specified. This last condition has led to the
almost exclusive use of parametric $SFR(t)$ histories.

Presently, methods
of comparing simulated C-M diagrams with observations can be classified
according to the statistical criterion used in the comparison. A few
examples of the variety in these categories are Tolstoy (1995) and
Mould et al. (1997) who use full maximum likelihood statistics,
Dolphin (1997) and Ng (1998) who use chi-squared statistics, and
Aparicio et al. (1997) and Hurley-Keller et al. (1998) who break the
C-M diagrams into luminosity functions before constructing the
statistical estimator.

We (Hernandez, Valls-Gabaud \& Gilmore 1999a,1999b;
Hernandez, Gilmore \& Valls-Gabaud 1999) have developed a
variational calculus method of solving directly for the maximum
likelihood $SFR(t)$, which does not require any assumptions on the
function one is trying to recover, or to evaluate the likelihood of
any of the $SFR(t)$'s being considered (all continuous functions of
time). We construct an integro-differential equation which is solved 
to find a $SFR(t)$ which yields a vanishing first variation for the
likelihood. At each iteration the $SFR(t)$ is solved with an arbitrary
time resolution. Conveniently, computation times scale only linearly
with this time resolution. This allows a very detailed reconstruction of
the $SFR(t)$, which would be prohibitively expensive in a parametric
decomposition of the $SFR(t)$.

Full details of the model, the extensive tests and calibrations, and
its application to new HST data for the Galactic dSph satellites, are
presented in the references noted (Hernandez et al. 1999, 1999a, 1999b). In
this paper we summarise the method and its validity, then simulate
GAIA observations of an old metal-poor stellar population, a young
metal rich population, and a mixed population. These simulations show
that GAIA data can indeed meet the scientific goal required, quantify
the metallicity accuracy needed, and quantify the photometric
precision required by GAIA at faint magnitudes.

\section{Deriving star formation histories}

Our goal is to recover the star formation history which gave rise to
an observed population of stars, described by $SFR(t)$, the star
formation rate as a function of time. For fully general applicability,
we shall assume nothing about the $SFR(t)$ we are trying to recover,
except that it should be a continuous function of time. One general
constraint will be the total number of stars produced, which provides 
the SFR normalisation condition, over the range of masses over
which stars can be observed. In the general case, we are concerned
only with that fraction of the total star formation which produced
stars still readily observable today, which corresponds in external
galaxies and at large distances in the Galaxy to masses greater than
about 0.5 solar masses. We emphasise that this mass constraint is
relevant only in that it must be lower than the oldest turnoff mass of
relevance to the application at hand. Also note that this requires an
accurate determination of the local completeness limit in any modelled
data set, but does not imply that this completeness limit have a value
near unity.

The final observed C-M diagrams as a function of place are the result
of the star formation histories in those places, later dynamical
evolution, and also of the relevant initial mass function, the
metallicity and the stellar evolutionary processes.  As we see later,
it is essential that the IMF near the turnoff, and the star by star
metallicity, be independently known. It is this requirement which is
the primary science case for GAIA to determine stellar metallicity,
and which constrains the requisite photometric performance. [Note that
there are other astrophysical systems for which this essentially
holds, and for which only the star formation history is poorly
known. Examples of such systems are some of the dwarf spheroidal
companions to our Galaxy, whose star formation histories we have
derived.]

In the Milky Way, we are primarily interested in  stars which are
observable at the distance of the Galactic centre, and the stellar
edge of the disk. This determines the mass regime over which the
initial mass function needs to be well established. Theoretical
studies of stellar isochrones have advanced significantly over the
last decade, and now there seems to be little uncertainty in the
physical properties of stars over the mass range 0.6--3 solar masses,
during all but the shortest lived evolutionary periods. Here we are using the
latest Padova isochrones (Fagotto et al. 1994, Girardi et al. 1996),
including most stages of stellar evolution up to the RGB phase.  Our
detailed inferences will depend on the precise details of the
isochrones we use. Our aim here is not to insist upon any
particular age calibration, and indeed
any and all isochrones can be used. The key feature of this method is
that it provides a robust and objective relative age ranking of the
stellar population. The mapping of that age ranking onto an external
time scale is isochrone-dependent, insofar as different isochrones do
not (yet) all agree, but the relative distributions are robust.

\subsection{The method}

Consider a fixed set of observations $A=(A_1,...,A_n)$, which 
result from a model which belongs to a certain known set of
models $B={B_1,...}$. We want to identify the model which has the highest
probability of generating the observed data set, $A$. That is, we
wish to identify the model which maximizes $P(AB_i)$, the joint
probability of $A$ occurring for a given model $B_i$. From the
definition of conditional probabilities,
\begin{equation}
P(AB_i)=P(A|B_i)\cdot P(B_i)=P(B_i|A) \cdot P(A)
\end{equation}
where $P(A|B_i)$ is the conditional probability of observing A given
that a specific
model $B_i$ occurred, $P(B_i|A)$ is the conditional probability
of model $B_i$ given the observed data $A$, and $P(A), P(B_i)$ are the
independent probabilities of $A$ and $B_i$, respectively. Further, if
the $B_{i}$s are exclusive and exhaustive,
\begin{equation}
P(A)=\int_{i} P(A|B_i) \cdot P(B_i)=1/C
\end{equation}
where $C$ is a constant, so that equation~(1) becomes:
\begin{equation}
P(B_i|A)=C\cdot P(A|B_i)\cdot P(B_i)
\end{equation}
which is Bayes' theorem. $P(B_i)$ is called the {\it prior}
distribution, and defines what is known about model $B_i$ without any
knowledge of the data. As we want to maximize the importance of the
data in our inference process, we adopt the hypothesis of equal prior
probabilities. Finding the maximum likelihood model under this
assumption is hence simplified to finding the model $B_i$ for which
$P(A|B_i)$ is maximized. Our set of models from which the optimum
$SFR(t)$ is to be chosen includes all continuous, twice differentiable
functions of time, with the external constraint 
that the total number of stars formed does not
conflict with the observed C-M diagram.

In order to find the $SFR(t)$ which maximizes the probability of the
observed C-M diagram resulting from it, we first have to introduce a
statistical model to calculate the probability of the data resulting
from a given $SFR(t)$. Consider one particular star, having an
observed luminosity and colour, $l_{i}, c_{i}$, and an intrinsic
luminosity and colour $L_{i}, C_{i}$, where the index $1<i<n$
distinguishes between the $n$ observed stars making up the C-M
diagram. The intrinsic and observed quantities will not be identical,
due to observational errors.  The probability of this observed point
being a star actually described by a particular isochrone $C(L;t_j)$,
i.e., being one of the stars formed at time $t_j$ as part of the rate
$SFR(t_j)$, will be given by:
\begin{equation}
P_{i}\left(t_{j}\right) = SFR(t_{j}) \; {\rho(L_i;t_{j}) \over{\sqrt {2
\pi} \; \sigma(l_i)}}  \; 
\exp\left(-\left[C(L_i;t_{j})-c_{i}\right]^2 \over {2 \; \sigma^2(l_i)}
\right)  
\end{equation}

In equation~(4) $\sigma(l_i)$ denotes the observational error in the
measurement of the colour of the $ith$ observed star, which is a
function of the luminosity of this star, and which we are assuming
follows a Gaussian distribution.  For simplicity here we
only consider errors in colour, which increase with decreasing
luminosity, in a way determined by the particular observation. In this
case $L_{i}=l_{i}$ which we adopt throughout, the generalization to an
error ellipsoid being trivial.  $C(L_i;t_j)$ is the colour the
observed star having luminosity $l_i$ would have if it had actually
formed at $t=t_j$.  $\rho(L_i;t_{j})$ is the density of stars along
the isochrone $C(L;t_j)$ around the luminosity of the observed star,
$l_i$, for an isochrone containing a unit total mass of
stars. Therefore, for stars in their main sequence phase,
$\rho(L;t_{j})$ is actually the initial mass function expressed in
terms of the luminosity of the stars. Further along the isochrone it
contains the initial mass function convolved with the appropriate
evolutionary track. Finally, $SFR(t_j)$ indicates the total mass of
stars contained in the isochrone in question, and is the only quantity
in equation~(4) which we ignore, given an observational C-M diagram, an
initial mass function and a continuous set of isochrones.

The probability of the observed point $l_{i}, c_{i}$ being the result
of a specific $SFR(t)$ will therefore be:

\begin{equation}
P_{i}\bigl( SFR(t)\bigr) = \int_{t_0} ^{t_1} SFR(t) \; G_{i}(t)\;  dt
\end{equation}
where
$$
G_{i}(t)= {\rho(L_i;t) \over{\sqrt{2 \pi}\; \sigma(l_i)}} \; 
\exp \left(-\left[C(L_i;t)-c_{i}\right]^2 \over {2 \; \sigma^2(l_i)} \right)
$$
and where $t_0$ and $t_1$ are a maximum and a minimum time needed to
be considered in a specific astrophysical or observational situation. 
We shall refer to $G_{i}(t)$
as the likelihood matrix. At this point we introduce the hypothesis
that the $n$ different observed points making up the total C-M diagram
are independent events, to construct:

\begin{equation}
{\cal L}= \prod_{i=1}^{n} \left( 
\int_{t_0} ^{t_1} SFR(t) \; G_{i}(t) \; dt \right)
\end{equation}
which is the probability that the full observed C-M diagram resulted
from a given $SFR(t)$. The discussion to here is essentially a review, 
and can be found similarly in, e.g., Tolstoy \& Saha (1996).

The remainder of the development is entirely new. We have used
equation (6) to construct the Euler equation of the problem, and hence
obtain an integro-differential equation directly for the maximum
likelihood $SFR(t)$, independent of {\it a priori} assumptions.  It is
the functional ${\cal L}(SFR(t))$ which we want to maximize with
respect to $SFR(t)$ to find the maximum likelihood star formation
history.

The condition that ${\cal L}(SFR)$ has an extremal can be written as
$$
\delta {\cal L}(SFR)=0,
$$
and the techniques of variational calculus brought to bear on the
problem. Firstly, we develop the product over $i$ using the chain rule
for the variational derivative, and divide the resulting sum by ${\cal
L}$ to obtain:

\begin{equation}
\sum_{i=1}^{n} \left(
{\delta \int_{t_0} ^{t_1} SFR(t) \, G_{i}(t) \, dt} \over {\int_{t_0}^{t_1}
SFR(t) \, G_{i}(t) \, dt} 
\right) =0
\end{equation}

In order to construct an integro-differential equation for $SFR(t)$ we
introduce the new variable $Y(t)$ defined as:

$$
Y(t)=\int{ \sqrt {SFR(t)} \, dt} \Longrightarrow  SFR(t)=\left( {dY(t)
\over dt} \right)^2
$$

Introducing the above expression into
equation~(7) and developing the Euler equation yields, 

\begin{equation}
{d^2 Y(t)\over dt^2}\sum_{i=1}^{n} \left( G_{i}(t) \over I(i)\right)
=-{dY(t)\over dt}\sum_{i=1}^{n} \left( dG_{i}/dt \over I(i)\right)
\end{equation}
where 
$$
I(i)=\int_{t_0}^{t_1} SFR(t) \, G_{i}(t) \, dt
$$

We have thus constructed an integro-differential equation whose
solution yields a $SFR(t)$ for which the likelihood has a vanishing
first variation.  This in effect has transformed the problem from one
of searching for a function which maximizes a product of integrals
(equation 6) to one of solving an integro-differential equation
(equation 8).  Solving equation~(8) will be the main problem, as this
would yield the required star formation history directly, without
having to calculate ${\cal L}$ explicitly over the whole functional space
containing all the possible $SFR(t)s$.

One may now implement an iterative scheme for solving equation(8), the
details of which are given in Hernandez, Valls-Gabaud \& Gilmore
(1999). Given the complexity of the isochrones, the initial mass
function and the unknown star formation histories we are trying to
recover, it is not possible to prove convergence analytically for the
implemented iterative method. Hernandez et al. show that the method
works remarkably well for a wide range of synthetic C-M diagrams
produced from known $SFR(t)$'s, independent of the initialisation
used.

\subsection{Numerical Implementation}

An important point of implementation, rather than general principle,
involves calculation of the isochrones.
To produce a
realistic C-M diagram from a proposed $SFR(t)$ requires firstly a
method of obtaining the colour and luminosity of a star of a given
mass and age. Interpolating between isochrones is a risky procedure
which can imprint spurious structure in the inference procedure, given
the almost discontinuous way in which stellar properties vary across
critical points along the isochrones, and how these critical points
vary with time and metallicity.  

To avoid this we use the latest Padova (Fagotto et al. 1994, Girardi
et al. 1996) full stellar tracks, calculated at fine variable time
intervals, and a careful interpolating method which uses only stars at
constant evolutionary phases to construct an isochrone library. We
calculate 100 isochrones containing 1000 uniformly spaced masses each,
with a linear spacing between 0.1 and 15 Gyr, which determines the
time resolution with which we implement the method to be 150 Myr. An
arbitrary time resolution can be achieved using a finer isochrone
grid, which  increases the calculation times only linearly with the
number of intervals. 

Having fixed the isochrones, we now need to specify the manner in which
the density of stars will vary along these isochrones, i.e. an IMF. 
We use the IMF derived by Kroupa et al. (1993), where a single fit to
this function is seen to hold for stars towards both Galactic poles,
and for all stars in the solar neighbourhood. In analyzing the stellar 
distribution towards the Galactic poles, a wide range of metallicities
and ages is sampled, and care was taken to account for all the
effects this introduces, including the changing mass-luminosity
relation at different ages and metallicities, completeness effects as a
function of luminosity and distance, and the contribution of binaries.
%At this point we shall assume their result to be of universal
%validity, and use their fit:
%
%\begin{equation}
%\rho(m) \propto \left\{
%\begin{array}{rl}
%     m^{-1.3} &  0.08M_{\odot} <m\le 0.5 M_{\odot} \\[1.0 ex] 
%     m^{-2.2} &  0.5M_{\odot} <m\le 1.0 M_{\odot} \\[1.0 ex]
%     m^{-2.7} &  1.0M_{\odot} < m
%\end{array}\right.
%\end{equation} 

We normalise the mass distribution such that a unit total mass is contained
upwards of $0.08 M_{\odot}$, although only stars in the mass range
$0.6 - 3 M_{\odot}$ can end up in the C-M diagram.
We can now choose a $SFR(t)$, and use the IMF  to populate our
isochrones and create a synthetic C-M diagram, after including
``observational'' errors, assumed to be Gaussian-distributed on log(T).

It is worth noting that the slope of the low-mass IMF, $-1.3$, is in
exact agreement with the one derived from very recent studies of metal-poor
globular clusters, and the UMi dwarf spheroidal galaxy, in both cases
from very deep HST data. It thus seems well-established.

\section{Testing the method: a summary}

To illustrate the validity of the method, we present here a small
subset of the simulation results of Hernandez et al., where further
details may be found. The method used here is to create an artificial
colour-magnitude diagram from a set of adopted star formation
histories, and then to apply the method above to deduce a star
formation history from the CMD. The true simulation input and the
derived output are compared, the insensitivity to assumed star
formation histories is apparent, the reliability of the method is
illustrated, while the extreme age-metallicity degeneracy which
irreducibly affects any analysis of stellar photometric data is explored.

\subsection{A simple two-burst example}

\begin{figure}
\epsfig{file=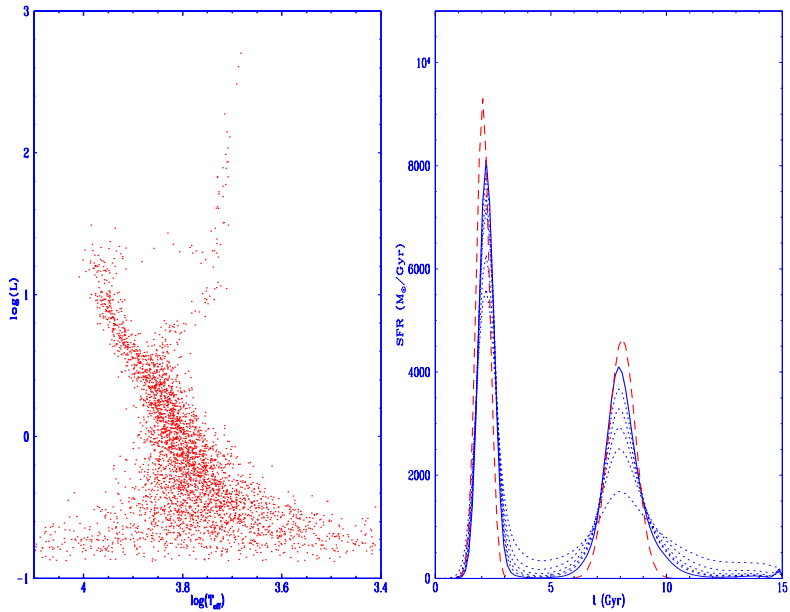,angle=0,width=11cm,height=9.0cm}
%\vspace{9.0truecm}
\caption[]{Left: Synthetic C-M diagram 
resulting from a simple two-burst input $SFR(t)$.
Right: Input $SFR(t)$, dashed line. Also shown
are the 3, 6, 9, 12 and 15 iterations of the inversion method, as dotted
curves.  The 20th iteration is given by the solid curve, showing
the convergence and a reliable reconstruction of the input $SFR(t)$.}
\end{figure}

As a first test we use a $SFR(t)$ consisting of two Gaussian bursts at
different epochs, of different amplitudes and total masses. This
$SFR(t)$ is shown by the dashed line in the right panel of Figure~(1),
where the time axis shows the age of the corresponding stellar
populations.  The left panel of Figure~(1) shows the resulting C-M
diagram which contains a total of 3819 stars.  A realistic
error structure was applied, from actual HST observations.

The calculation proceeds using the stars with
$\log(L)>0$ in Figure~(1), from which  we construct the matrix
$G_{i}(t)$. Since the colour of a
star having a given luminosity can sometimes be a multi-valued
function, in practice we check along a given isochrone, to find all
possible masses a given observed star might have as a function of
time, and add all contributions (mostly 1, sometimes 2 and
occasionally 3) in the same $G_{i}(t)$. Calculating this matrix is the
only slow part of the procedure, and is equivalent to calculating the
likelihood of one model. The likelihood matrix $G_{i}(t)$ is the only
input required by the numerical implementation. 
The total number of stars is used as a
normalisation constraint at each iteration, needed to recover $SFR(t)$
from $Y(t)$. As mentioned earlier, it is not necessary to calculate
the likelihood over the solution space being considered,
i.e. $G_{i}(t)$ is only calculated once, which makes the method highly
efficient.

In Figure~(1) we also show the results of the first 12 iterations of
the method every 2 iterations, which form a sequence of increasing
resemblance to the input $SFR(t)$. The distance between successive
iterations decreases monotonically at all ages, which together with
the fact that after 12 iterations no further change is seen, shows the
convergence of the method for this case. From the 2nd iteration
(lowest dotted curve in the burst regions) it can be seen that the
iteration of the variational calculus equation constructed from
maximizing the likelihood is able to recover the input $SFR(t)$
efficiently. The positions, shapes and relative masses of the two
bursts were correctly inferred by the 2nd iteration, although it took
longer for the method to eliminate the populations outside of the two
input bursts. The convergence solution is in remarkable agreement with
the input $SFR(t)$, and only differs slightly, as seen from
Figure~(1). No information was used in the inverting procedure beyond
that which is available from the synthetic C-M diagram, which was used
extensively in constructing the likelihood matrix $G_{i}(t)$, which is
the only input required by the inversion. The variational calculus
method recovers a $SFR(t)$ for which the first variation of the
likelihood vanishes, without assuming any {\it a priori} condition on
the $SFR(t)$, beyond being a continuous twice differentiable function
of time.

\begin{figure}
\epsfig{file=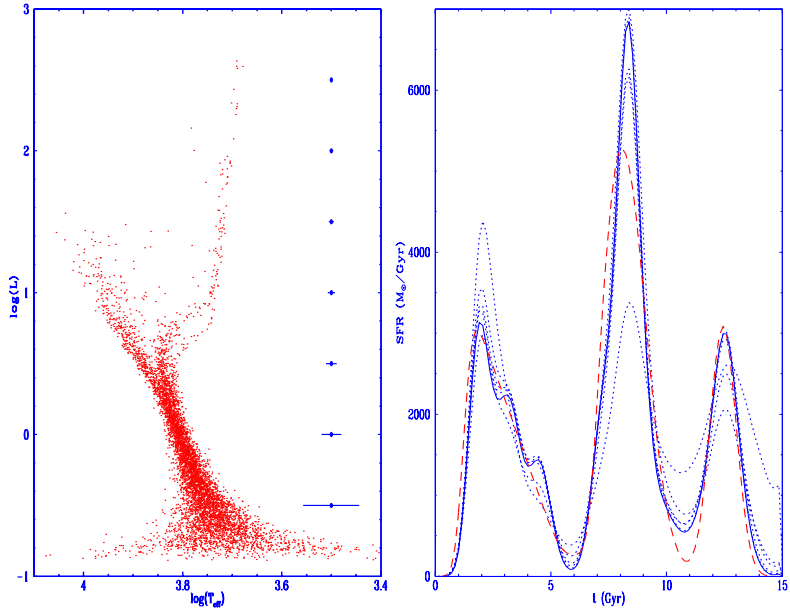,angle=0,width=11cm,height=9.0cm}
%\vspace{9.0truecm}
\caption{Left: A synthetic C-M diagram with smaller photometric errors.
Right: The input $SFR(t)$ is shown as a dashed line. Also shown
are the 2, 4, 6, 8 and 9 iterations of the inversion method, dotted
curves.  The 10th iteration is given by the solid curve, showing rapid
convergence and a good recovery of the input $SFR(t)$. }
\end{figure}

\subsection{Very old populations: sensitivity to photometric errors}

Given that the method is now seen to work, how well can it work? The
hardest case is for very old stars, where the colour-sensitivity to
age differences is small. For this experiemnt we consider data of
somewhat higher photometric precision that are routinely obtained with
HST, but are similar to those from large ground-based and future space
CCD arrays. Figure (2) has the same structure as Figure~1 above. The 
data quality
is reflected in the clearer C-M diagram, where the older population is
distinguishable from the noise of the younger main sequence.  The
right panel in Figure (2) shows the result of the inversion procedure,
and the rapidity of convergence, with only 10 iterations needed.  The
few stars in the oldest component which can be separated from the
younger main sequence are sufficient to accurately recover the shape
for this burst. That is, the method is able to deduce all the
information present in the data.

\subsection{Sensitivity to uncertainties in IMF, metallicity and binaries}

Uncertainties in the IMF, metallicity and binaries differ from simple
sample size or photometric error in inducing a systematic mismatch
between the isochrones used in any specific calculation and those
which describe the astrophysics of the C-M diagram being inverted.  
Of these effects, by far the most significant is the well known
age-metallicity degeneracy, which implies that a  reliable knowledge of
the abundance distribution of the stars in the sample is essential for
any reliable determination of age from photometric data.

\begin{figure}
\epsfig{file=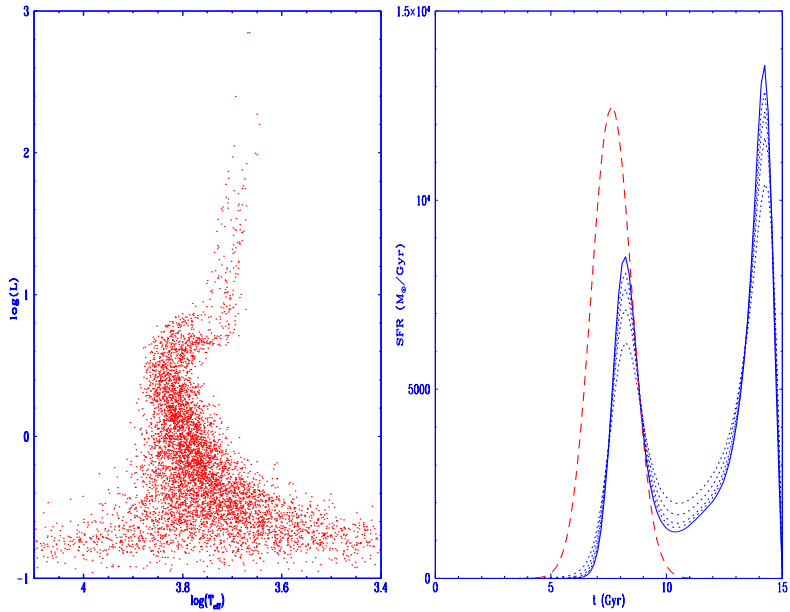,angle=0,width=11cm,height=9.0cm}
%\vspace{9.0 truecm}
\caption{Left: Synthetic C-M diagram 
produced using a metallicity
of $[Fe/H]=-1.7$ for the stars older than 7.5 Gyr and of
$[Fe/H]=-0.7$ for the stars younger than 7.5 Gyr.
Right: The input $SFR(t)$ is shown as the  dashed line. Also shown
are the  3, 6, 9, 12 and 15 iterations of the inversion method (dotted curves)
assuming a (wrong) 
constant metallicity ($[Fe/H]=-1.7$) for the entire evolution, which
confuses the method. 
The 18th iteration is given by the solid curve. 
}
\end{figure}

We therefore consider the following test, where the C-M diagram is
produced using a range of metallicities, and inverted assuming a
single metallicity. This is presented in Figure (3) with a
 C-M diagram which results from a gaussian $SFR(t)$,
where 
metallicity is not a delta function. In this case we assigned a metallicity of
$[Fe/H]=-1.7$ to stars older than 7.5 Gyr, and  $[Fe/H]=-0.7$
to stars younger than 7.5 Gyr, i.e., a crude enrichment
history. This is clearly seen in Figure (3, LHS), where the two
populations having different metallicities are evident, from the width
of the RGB. The noise level is the same as in Figure ~(1).
The result of applying the inversion method assuming a
single metallicity of $[Fe/H]=-1.7$ is shown in the right panel of
Figure (3).  The method correctly identifies the half of the $SFR(t)$
with the lower metallicity; the higher metallicity population is
totally misinterpreted. Actually, the age the inversion procedure
should assign to the high metallicity component is in fact greater
than 15 Gyr, which is in contradiction with the fixed boundary
condition of $SFR(15)=0$. This makes the inversion procedure somewhat
unstable, which in principle can be used to indicate that the
isochrones being used in the inversion procedure do not correspond to
the studied stars.  The two distinct giant branches seen in this C-M
diagram indicate a difference in the metallicities of both
populations. 

As it might have been expected, uncertainties in the metallicity
distort the inference procedure significantly, making the determination of
star formation histories robust only in cases where individual
metallicities are available.

\section{Conclusions concerning the method}

We can summarize our methodological results as follows:

1) We have introduced a variational calculus scheme for solving
maximum likelihood problems, and tested it successfully in the
particular case of inverting C-M diagrams.

2) Assuming a known IMF and metallicity we have presented a 
non-parametric method for inverting C-M diagrams which yields good
results when recovering stellar populations younger than 10 Gyr, with
data quality similar to those attained in current HST observations of
dSph galaxies.  Populations older than 10 Gyr can only be recovered
equally well from C-M diagrams with much reduced observational errors,
compared to current HST performance.

3) Uncertainties in the IMF and binary fractions result in
normalisation errors on the total $SFR(t)$. Given the existence of an
age-metallicity degeneracy on the colours and magnitudes of stars, an
error in the assumed metallicity results in a seriously mistaken
$SFR(t)$. This makes the version of the variational calculus approach
we present here useful only in cases where the metallicity of the
stars is knowable independently of the form of the colour-magnitude
diagram near the turnoff.

4) The main-sequence turnoff of the oldest stellar populations
corresponds to apparent magnitude $\sim 20$ at the Galactic centre,
and near the apparent outer edge of the disk. GAIA can therefore
determine the full star formation history of the near half of the
Milky Way for all ages. This provides a sufficiently large sample that
the global star formation history of our Galaxy can indeed be
determined by GAIA.
The star formation history at more recent times can
also be determined for larger distances, especially in the Magellanic
Clouds and the Sgr dSph. Such determinations however require metallicity
measures accurate to about 0.2 dex, and photometric data able to provide
stellar effective temperatures good to about 10\% of T$_{\rm eff}$, for
stars at and above the turnoff.

\end{article}


\begin{thebibliography}{}

\bibitem{}Aparicio A., Bertelli G., Chiosi C., Garcia-Pelayo J.M.,
1990, A\&A 240, 262 
\bibitem{}Aparicio A., Gallart C., 1995, AJ, 110, 2105
\bibitem{}Aparicio A., Gallart C., Bertelli G., 1997, AJ, 114, 669
\bibitem{}Chiosi C., Bertelli G., Meylan G., Ortolani S., 1989, A\&A, 219, 167
\bibitem{}Dolphin A., 1997, New Astronomy, 2, 397
\bibitem{}Fagotto F., Bressan A., Bertelli G., Chiosi C., 1994, A\&AS, 104, 365
\bibitem{}Girardi L., Bressan A., Chiosi C., Bertelli G., Nasi E., 1996, 
A\&AS 117, 113 
\bibitem{}Hernandez, X., Valls-Gabaud, D., \& Gilmore, G., 1999 MNRAS
304, 705
\bibitem{}Hernandez, X., Gilmore, G. \& Valls-Gabaud, D., 1999 MNRAS
in press
\bibitem{}Hernandez, X., Valls-Gabaud, D., \& Gilmore, G., 1999 MNRAS
submitted
\bibitem{}Hurley-Keller D., Mateo M., Nemec J., 1998, AJ, 115, 1840
\bibitem{}Kroupa P., Tout C.A., Gilmore G., 1993, MNRAS, 262, 545
\bibitem{}Mighell K.J., Butcher H.R., 1992, A\&A, 255, 26
\bibitem{}Mighell K.J., 1997, AJ, 114, 1458
\bibitem{}Mould J.R., Han M., Stetson P.B., et al., 
 %Gibson B., Graham J.A., Huchra J., Madore B., Rawson D.,
 1997, ApJ, 483, L41
\bibitem{}Ng Y.K., 1998, A\&AS, 132, 133
\bibitem{}Smecker-Hane T.A., Stetson P.B., Hesser J.E., Lehnert M.D.,
1994, AJ, 108, 507 
\bibitem{}Tolstoy E., 1995, PhD. Thesis, Groeningen University, The
Netherlands. 
\bibitem{}Tolstoy E., Saha A., 1996, ApJ, 462, 672

\end{thebibliography}
\end{document}